\documentclass[preprint,aps,nofootinbib,superscriptaddress]{revtex4}
\newcommand{\beq}{\begin{equation}}
\newcommand{\feq}{\end{equation}}

\begin{document}

\title{On Black Hole Scalar Hair in Asymptotically Anti de Sitter Spacetimes}
\author{Daniel Sudarsky} 
\email{sudarsky@nuclecu.unam.mx}
\affiliation{ Instituto de Ciencias Nucleares\\
Universidad Nacional Aut\'onoma de M\'exico\\
A. Postal 70-543, M\'exico D.F. 04510, M\'exico.\\}
\affiliation{Center for Gravitational Physics and Geometry,\\
Department of Physics, Penn State University,\\
University Park,PA 16802, USA.\\}
\author{Jos\'e Antonio Gonz\'alez}
\email{cervera@nuclecu.unam.mx}
\affiliation{ Instituto de Ciencias Nucleares\\
Universidad Nacional Aut\'onoma de M\'exico\\
A. Postal 70-543, M\'exico D.F. 04510, M\'exico.\\}
\begin{abstract}
 The unexpected discovery of  hairy black hole solutions in
 theories with  scalar fields simply by
considering asymptotically Anti de-Sitter, rather than asymptotically flat,
boundary conditions
is analyzed in a way that exhibits in a clear manner the differences between
the two situations.
 It is shown that the trivial Schwarzschild Anti de Sitter becomes
unstable in some of these
 situations, and the possible relevance of this fact for
the ADS/CFT conjecture is pointed out.

\noindent {\bf Key words:} {Black Holes, Hair, Anti de Sitter}
\end{abstract}
\maketitle

\section{Introduction}
The falsehood  of the no-hair conjecture for stationary black holes is
hardly  even disputed these days
as the list of  counterexamples becomes ever larger:  Einstein-Yang-Mills
\cite{EYM}, Einstein-Skyrme
\cite{ES}, Einstein-Yang-Mills-dilaton \cite{EYMd},
Einstein-Yang-Mills-Higgs \cite{EYMH}, and
Einstein-non-abelian-Procca \cite{EYMH} fields.  In some  sub-communities
the idea seems to be holding out
that a modified version  that applies only to stable black holes, could
remain valid, despite the fact
that for some of the examples above some claims of stable non-trivial
solutions exist in the literature.

One  place where  it seemed for a while that there was hope for a
restricted form of the conjecture was
the scalar field arena.  Here we had  the original no hair theorems of
Bekenstein
\cite{Bek0} covering
the case of  minimally coupled scalar fields with  convex potentials, other
theorems dealing with the case of
minimally coupled fields with arbitrary potentials were obtained
in \cite{Heusler} and  \cite{EH}.
The  so called  Bronnikov-Melnikov-Bocharova-Bekenstein
(BMBB) black hole ``solution"
\cite{bekbh} which corresponds to a spherical symmetric extremal black hole
 with
a scalar field  conformally coupled   to gravity seemed to represent a
discrete example of scalar hair,
as it was shown  \cite{XZ} that there are no other static spherically
symmetric  Black Hole solutions in
this theory.  Later on, it was shown  that this configuration, which
presents a divergence of
the scalar field at the horizon, cannot be considered as a regular black
hole solution  because the energy
momentum tensor is  ill-defined at the horizon \cite{sudzann}. Finally it
has been shown that
if one demands that the scalar field  be bounded throughout  the static
region then, there are no solutions  at all.
\cite{Zannias}.

 For more general cases of  non minimal coupling there are results \cite{Saa}
showing  that  under the assumption  that certain `` conformal factor"
doesn't vanish or blow up,  there
are no nontrivial black hole solutions.
 Next, there is a result by \cite{Bekmayo}  that  does not rely  on such
assumption, and
 which considers the existence of static , spherically symmetric black hole
solutions
in theories  in  which  the sign of the non-minimal coupling constant is
negative ( the only case not
covered by other theorems)  case. There it is shown that under certain
suppositions about the form of the
energy-momentum flux, there are no nontrivial solutions.
 In \cite{Igor} it is argued that these suppositions are not fully 
justifiable and numerical evidence is given against the
existence of such black hole that doesn't rely on these assumptions.

 It is therefore a rather unexpected development that hairy black hole
solutions have now been found in
both theories with minimal \cite{Min} as well as non-minimal \cite{Nonmin}
coupled scalar fields simply by
considering asymptotically anti de-Sitter, rather than asymptotically flat,
boundary conditions.
 Moreover these papers have strong indications that, under certain
conditions, the new solutions are stable.

 The purpose of this paper is to analyze the situation  regarding the
asymptotically  anti de Sitter
case, in light of  existing  results for the asymptotically flat case,
discuss the  points where the
differences are relevant and give a simple explanation of some of the
features of the new solutions and
point to some surprising conjectures that can be directly infer from this
understanding.  The method used in
this part is a  generalization of  one that
was successfully employed  in deriving a general characterization of hairy
black holes in a wide range of theories \cite{us}.

An added reason for interest in the asymptotically Anti-de-Sitter case, and
one we briefly touch in this paper is the  important place such
spacetimes have acquired in view of the AdS/CFT conjecture. In fact we will
argue that the results and conjectures that are pointed to in this
work, indicate a   difficulty for the notion that the ADF/CFT idea can have
as general  a validity as is normally deem to have.

\section{  Scalar Hair and Asymptotic conditions }
\label{sec2}
 We will  restrict consideration to the minimally coupled case as the
emergence of hair doesn't rely at all
on the more complicated non-minimal couplings. Thus we will consider a
theory  given by the action:

\beq
S=\int \sqrt{-g}d^4x \left[ \frac{1}{16\pi}(R-2\Lambda) - 1/2 (\nabla
\phi)^2- V(\phi) \right]
\feq
 where $\phi$ is a scalar field and  $V$ its potential,  $R$ is the scalar
curvature,  and  $\Lambda $ is the true cosmological constant
( by which we mean that the minimum of the scalar potential has been set to
$0$, and any nonzero part has
 been absorbed into $\Lambda $).  Now we
restrict attention to static spherically symmetric regular black holes
whose exterior we parameterize as
\beq
ds^2 = -e^{-2\,\delta}\mu\,dt^2 +\mu^{-1}\,dr^2 +r^2\,d\Omega^2 \ ,
\label{eq:lel}
\feq
 where $\mu$ and $\delta$ are functions of $r$. Note that  $ \delta $ can
be taught of as representing an
 additional red-shift, beyond the one
that could be inferred from the geometry of the static hypersurfaces (i.e.
those that are normal to the Killing Field).
 The condition for a presence of a regular horizon at $r_H$ requires the
vanishing of $\mu$ there.
 It is customary to parameterize $\mu$ as
\beq
\mu(r) = 1- \frac{2m(r)}{r} + \lambda r^2
\label{def:mu}
\feq
 where the asymptotic geometry is controlled by the parameter $\lambda$
(i.e $\lambda =0$ for the asymptotically
  flat case,  $\lambda >0$ for
the asymptotically Anti de Sitter case, and $\lambda < 0$ for
the asymptotically  de Sitter case).  Einstein's equations give in this case
\beq
\mu^\prime=8\,\pi\,r\,{T^t}_t + {{1-\mu}\over r},\label{eq:mu}\qquad
\delta^\prime={{4\,\pi\,r}\over{\mu}}\,({T^t}_t-{T^r}_r ),\label{eq:delta}
\feq
where prime stands for differentiation with respect to $r$. The scalar
field equation  can be written as;
\beq
\mu \phi'' + [ (1/r) ( \mu +1)  + 4 \pi r ({T^t}_t+{T^r}_r )  ]\phi' -
{{\partial V} \over {\partial \phi}} =0
\label{eq:phi}
\feq

We must point out that in the above formulas refer to the ``total energy 
momentum" tensor $T_{\mu \nu}$ which is related to the energy momentum of 
the scalar field
 $T_{\mu \nu}(\phi)$  as  $ T_{\mu \nu} =T_{\mu \nu}(\phi)- g_{\mu \nu} 
\Lambda /
(8\pi) $ .

The main tool of our analysis is simply the conservation for the $r$
component of the total energy
momentum tensor $
{T^\mu}_{r;\mu}=0 $,
 which,  through the use of Einstein's equations can be written as \cite{us}:
\beq
e^\delta\,(e^{-\delta}\,{T^r}_r)^\prime=
{1\over{2\,\mu r}}\left[ ({T^t}_t-{T^r}_r)
 + \mu (2T -3  {T^t}_t -5{T^r}_r)\right],\label{eq:teo}
\feq
where $T$ stands for the trace of the stress energy tensor.

 The  energy momentum tensor of the scalar field by itself satisfies then:

\beq
e^\delta\,(e^{-\delta}\,{T (\phi)^r}_r)^\prime=
{1\over{2\,\mu r}}\left[ (1 + r^2 (-\Lambda) ) ({T(\phi)^t}_t-{T(\phi)^r}_r)
 + \mu (2T (\phi) -3  {T(\phi)^t}_t -5{T(\phi)^r}_r)\right]
\label{teo2}
\feq

 Here we can review the reasons behind the fact that there is  no
nontrivial scalar field   in the exterior of such
black holes in the asymptotically flat case with
$\Lambda =0 $.  First the regularity at the horizon  requires that mixed
components $ T (\phi)^{\mu}_{\nu} $ must be
bounded at the horizon since the scalar $ T (\phi)^{\mu}_{\nu}{T
(\phi)^\nu}_{\mu}$ is in this case a sum of
non-negative terms. Next we note that the vanishing of $\mu $ at the
horizon indicates that on the horizon
$T (\phi)^r_r=T (\phi)^t_t $ which is negative  definite as follows from
the Weak Energy Condition (WEC)  which is satisfied in particular by
minimally coupled  scalar fields ( in fact, in our case  we have $T
(\phi)^r_r = 1/2 \mu (\phi')^2 - V $ and
$T (\phi) ^t_t = T (\phi) ^{\theta}_{\theta} =T (\phi) ^{\varphi}_{\varphi}
= - 1/2
\mu (\phi')^2 - V $ ).  Next,  it follows from the WEC  that
$({T^t}_t-{T^r}_r) <0$ and from  fact that for the situation at hand the
combination $(2T (\phi) -3  {T(\phi)^t}_t -5{T(\phi)^r}_r)  $ is $-3 \mu
(\phi')^2$ and thus  non-positive, that   the left hand side of eq.
(\ref{teo2}) is non positive, and thus  that $e^{-\delta}{T^r}_r (\phi)
$ is a  decreasing function of $r$. It is thus impossible for this function
to approach zero as it would be required from asymptotic
flatness,   the boundary condition that is relevant in   this case.
 The point is that if we consider  now asymptotically  Anti de Sitter
boundary conditions, and a  negative  cosmological
constant  two aspects of the above discussion remain unchanged : $e^{-\delta}
T(\phi)^r_r  $ is negative  definite at the horizon, and it is a decreasing
function of $r$.  Thus the reasons behind the
possibility of scalar black hole hair in  such case is the fact that in the
Anti de Sitter case one can allow $T (\phi)^r_r
$ to go to a finite (and negative) constant value at infinity which results in an
effective cosmological constant which differs from the true
cosmological constant of the theory. In fact we can now restate the  result
of the above analysis for the asymptotically Anti de Sitter case as follows:  

{\bf Theorem.} 
{\it There are no nontrivial static and spherically symmetric black 
hole solutions in asymptotically Anti-de-Sitter case  in which the
asymptotic behavior corresponds to the Anti-de-Sitter spacetime
with the true cosmological constant}. 

In other words, the asymptotically
Anti de Sitter region correspond to one where the effective
cosmological constant is :

\beq
 \Lambda_{eff} = \Lambda + 8\pi  V( \phi_\infty)
\feq

This is in fact in essence the difference between the asymptotically flat, 
$\Lambda =0$ case vs. the asymptotically Anti de Sitter, $\Lambda \not =0$ case, 
i.e. the fact that in the former case we must require $V$ to go to $0$ at infinity,
while in the latter case any nonzero asymptotic value of $V$ can be  absorbed 
into the effective cosmological constant.
The theorem above in fact ensures that such asymptotic value
is in fact nonzero and that such absorption can not be done
without.
Note that for a nontrivial black hole $V(\phi_\infty) > V(\phi_{r_H})\geq 0$, 
thus $\Lambda_{eff}> \Lambda $ and  the asymptotic behavior of the 
spacetime in then less " Anti de Sitter like" than would
have been expected from the  actual value of the  true cosmological
constant.

 Moreover, as one is interested in situations in which the scalar field
converges to a finite value at infinity we can look at the scalar field
equations and note that  a  necessary condition for such behavior is that
the field should go to an extremum of the potential at infinity.
Thus our general result, that $ T (\phi)^r_r $ must be a decreasing
function, together with the fact that in this regime it coincides with $V$,
suggests that the extremum of $V$ must be approached from below at
infinity, and thus, that such extremum must be a maxima.  
In fact, assuming that the scalar field converges to a finite limit $\phi_\infty$ at
infinity, and that at this point the potential takes a finite value, and
we write the asymptotic solution as $ \phi = \phi_\infty + f(r)$, with
$f\to 0 $ at $\infty$.
The asymptotic form of the scalar field equation is
 \beq
\lambda  r^2 f''+ 4\lambda r f' -  {{\partial
V} \over {\partial \phi}}|_{\phi_\infty} - {{\partial^2
V} \over {\partial \phi^2}}|_{\phi_\infty}  f =0
\label{eq:phi2}
\feq
From here we see that ${{\partial V} \over {\partial \phi}}|_{\phi_\infty} =0$.
The  solution of this equation that goes to zero at infinity is of the 
form $f = 1 / r^{\beta} $ with $\beta>0$.
 Substituting in eq. (\ref{eq:phi2}) one concludes that 
\beq
\beta = {{3}\over{2}} \pm \sqrt{{{9}\over{4}} + {{1}\over{\lambda}} {{\partial^2V} \over {\partial \phi^2}}|_{\phi_\infty}}.
\label{beta}
\feq
On the other hand, from the Einstein equations we have that $m' \sim r^4 {f'}^2$ so
the convergence of $m$ requires that $\Re(\beta)>3/2$ and thus the type of
oscillating behavior suggested in \cite{Min} can not occur.

If ${{\partial^2V} \over {\partial \phi^2}}|_{\phi_\infty} > 0$ one of the roots in 
(\ref{beta}) makes $\phi$ divergent, requiring a fine-tunning to avoid this divergence. 
So, although it is not possible to rule out a solution in this case, we are going to
consider the case in which the scalar field goes to a local maximum at infinity, i.e.
$0>{{\partial^2V} \over {\partial \phi^2}}|_{\phi_\infty} > -{{9}\over{4}}\lambda$;
which are in fact the cases in which solutions have been found.

 There are several interesting points that come out of this analysis: First
we note that  we can chose the cosmological constant
to be such  that the effective cosmological constant vanishes, leading to
scalar field hair for  black holes in the asymptotically
flat context! ( the price paid for this possibility  is the introduction of
a fine tuned  cosmological constant). The next point concerns
the issue of stability. This, as already mentioned has been considered an
important hope to salvage something of the no hair
conjecture. The point is that in these theories the Schwarzschild -Anti de
Sitter black holes are also trivial solutions, and thus, one
could hope  that if,
 as indicated by the available  evidence (See \cite{Min}),  the new,
nontrivial black hole solutions are stable, there would seem  be
a clear violation of even the weakened version ( i.e. the version dealing
solely with stable black holes) of the no hair conjecture.
 The first issue that comes to mind is what is the meaning of stability in
the asymptotically Anti- de -Sitter context. Normally, what
one means by stability in principle, is the following: Given initial data
corresponding to the configuration in question, are there
small perturbations of these, that grow without bounds with evolution in
time?. The point is that,  as  the  Anti de Sitter spacetime
is not globally hyperbolic (i.e. has no
Cauchy hypersurfaces) there is in  principle no well posed initial value
formulation that would allow one to  investigate such
question so there is no possible  meaningful answer to it, and thus no
meaning to the question. The only way to go around this problem
seems to be to fix ``boundary conditions at infinity" throughout the time
of evolution so as to obtain a well posed initial
value problem. Then the issue of stability relates to situations in which
we consider small perturbations as in the previous
discussion but keep among other things the value of the scalar field fixed
at infinity. It is in this regard that the new black hole
solutions could possibly be stable. We will assume from here on that such
stability is in fact verified for these solutions. Now let us ask
ourselves whether such stability is indeed surprising or not. The first
thing we note is that, as mentioned before,  at infinity the scalar
field is sitting at a local maximum  of the potential, and thus, that the
stability is intimately connected with the fact
that we are dealing with a problem of evolution with fixed conditions at
infinity, for otherwise our intuition tell us that under
perturbations the  field would tend to roll  down the potential.

 What lies behind the stability of the new stable configurations ought to
be then, that they represent the configurations of ``minimal mass"  (See
\cite{Mass} and references therein for a formal definition of mass in this
context and comparison with alternative ones)
among those that have a given black hole area\footnote{We are assuming a 
generalization of the ideas presented in \cite{Sud-Wald}.} 
and fixed value of the scalar field at infinity.  Assuming that  this is  the case,
the following conjecture naturally follows:
 For such situations in which the nontrivial  black hole is stable,
 the trivial solution with similar boundary condition  i.e. the standard
Schwarzschild Anti-de-Sitter solution  with the same boundary conditions
(with the
 scalar field frozen at the top of the potential throughout
spacetime) should be  unstable. In particular  we  can consider the
situation in which a
 fine tuning has made the effective cosmological constant
equal to zero,   and then, the plain old Schwarzschild solution should be
unstable.
This situation  is analogous to the case of the  magnetically
charged Reisner Nostrom Solution which is stable within Einstein Maxwell
Theory, but is unstable within Einstein Yang Mills Theory
\cite{unstableRN}.

 Finally, we note that in \cite{Min},  stable as well  as unstable
nontrivial solutions were found.
What lies behind the difference in these situations?.  It is natural to
assume it has to do with a change in the sign of the mass difference
between the two solutions with the same horizon area and asymptotic  value
of the scalar field. We note that in the situation at hand  such
difference can  have either sign depending on the details of the scalar
potential. In fact let's compare
 $M_2 (r_H)$,  the mass of a nontrivial  static black hole of
 radius $r_H$ with $M_1(r_H)$ the mass of  the corresponding Schwarzschild
-Anti de
Sitter black hole of the same radius (by  black hole radius we mean 
$r_H=\sqrt{A/4\pi}$ where $A$ is the horizon area.
And by mass we mean the Hamiltonian mass as defined in \cite{Mass}, which  in 
the present situation can be evaluated simply  by   $M=\lim_{r\to \infty} 
m(r) $).

In the latter case the solution is just given by setting
$\phi\equiv\phi_\infty$, $\delta \equiv 0$ and $ \mu(r) = 1- 2M_1 /r +
\lambda r^2$
 with  $M_1$ the corresponding  mass of the black hole of radius $r_H$ so
\beq
M_1 (r_H) = \frac{r_H}{2} ( 1 + \lambda r_H^2)
\feq
 In the case of the nontrivial black hole the mass is obtained  from the
equation for $m'$ that follows from  eqs. (\ref{eq:mu}) and
(\ref{def:mu});

\beq
m' = - 4\pi r^2 T^t_t + (3/2)\lambda r^2 = (r^2 / 2) [ (3\lambda +\Lambda)
+ 8\pi V(\phi) + 4\pi \mu  (\phi')^2]
\label{eq:m'}
\feq
 First, we note that the requirement that $m'\to 0$ at $\infty$ implies that
\beq
 \lambda =- (1/3) [\Lambda  +8 \pi V(\phi_\infty)]= -1/3 \Lambda_{eff}.
 \label{eq:l}
\feq
 Next the vanishing of $\mu$ at the horizon requires that $m(r_H)= r_H/ 2 (
1 + \lambda r_H^2) =M_1(r_H)$ , so  we can write,
 using eq. (\ref{eq:l}) the
mass of the  nontrivial black hole as:
\beq
M_2 = m(r_H)  +  4 \pi \int_{r_H}^{\infty}  r^2 [  V(\phi)- V(\phi_\infty)
+ (1/2) \mu  (\phi')^2] dr
\feq
Thus the sign of $M_2 - M_1$ depends on the integral which could have
either sign depending on the details of
 the potential and the horizon radius.
Note that this is in contrast with the situation that arises, say, in
Einstein Yang Mills
 Theory and its hairy black holes in the asymptotically
flat context, where the mass of the nontrivial black hole is

 \beq
M_2 = m(r_H)  +  4 \pi \int_{r_H}^{\infty}  r^2 [  (1/r^2) V(w) + (1/2) \mu
(w')^2]dr
\feq

 where $w$ parameterizes the Yang Mills field (as in \cite{EYM}) and  the
term  $V(w) = (1/2)(1-w^2)^2$ which arises from the self interaction of
the non-abelian fields plays a role of an effective -- and non-negative--
potential in this situation). It is clear
 that in this case the mass of the
hairy black hole is larger than that of the Schwarzschild solution with the
same horizon area.
In fact it should be  quite easy to numerically test whether the change in
stability is associated with the change in the
sign of this integral and we expect to do this in the near future.

 Finally a note regarding the No Hair Conjecture and the nature of the
counterexample obtained in \cite {Min}. The  standard understanding is that
one  says that  there are hairy black holes, if within a specific theory,
the boundary conditions and charges at infinity are
not sufficient to uniquely specify a stationary black hole solution. If one
were to take the position (not advocated by these authors, but
apparently advocated by the authors of \cite{Min}) that one only considers
stable black holes in this context, then in order to say that one has
found hair  it is not enough to show that the new solutions are stable, one
must also ensure that the trivial solution remains
stable as well. 
In fact using the result of the analysis of \cite{Inestability}, 
we can easily prove that for certain values of the parameters
the Schwarzschild Anti de Sitter solution will be unstable in this context, 
and thus the issue of the violation of the weakened version of the No Hair Conjecture
in the scalar field arena, would be far from settled.

The perturbations around the Schwarzschild Anti de Sitter black hole with
$\phi \equiv \phi_{\infty}$ are described by:
\beq
\begin{array}{c}
\mu(t,r)=\mu_{0}(r)+\epsilon\mu_{1}(t,r) \\
\delta(t,r)=\epsilon\delta_{1}(t,r) \\
\phi(t,r)=\phi_{\infty}+\epsilon\phi_{1}(t,r) \\
\frac{\partial V}{\partial \phi}=-\epsilon\phi_{1}(t,r)\alpha^2
\end{array}
\feq
where $\alpha^2=-\frac{\partial^2 V}{\partial \phi^2}|_{\phi_{\infty}}$ and 
$\mu_{0}(r)=1-\frac{2M}{r}+\lambda r^2$.
The first order perturbated equation for the scalar field is:
\beq
\ddot{\phi_1}=\mu_0\left[\mu_0\phi_1^{\prime\prime} + \left(\frac{2}{r}\mu_0 
+ \mu_0^{\prime}\right)\phi_1^{\prime} + \alpha^2\phi_1 \right]
\label{pert}
\feq
and the first order perturbed Einstein equations are identically satisfied by 
$\delta_1 = 0$ and $\mu_1 = 0$.

Equation (\ref{pert}) can be written as $\ddot{\phi_1}=-A\phi_1$ where
$A = -D^{a}D_{a} + V$ and $D_{a}$ is the covariant derivative associated with the 
auxiliary spatial metric
\beq
^{(3)}ds^2 = dx^2 + r(x)^2(d\theta^2+{\rm sin}^2\theta d\varphi^2)
\feq 
where
\beq
x(r) = \int^{r}_{r_H}dr^{\prime}\left(1-\frac{2m}{r^{\prime}}+\lambda r^{\prime 2}\right)^{-1}.
\feq 
Note that when $r \rightarrow \infty$, $x$ converges to a finite constant that we denote by $c$.
In this way we can write:
\beq
D^{a}D_{a}=\frac{d^2}{dx^2}+\frac{2\mu_0(x)}{r(x)}\frac{d}{dx}.
\feq
As mentioned in \cite{Inestability}, if $\psi$ is a vector of the Hilbert space 
$L^2$ with inner product 
\beq
\langle\psi_1,\psi_2\rangle =\int^{2\pi}_{0}\int^{\pi}_{0}\int^{c}_{0} r^2 \psi_1\psi_2 {\rm sin}\theta dx d\theta d\varphi
\feq
such that $\langle\psi,A\psi\rangle < 0$ then, the configuration is unstable. 

If we choose $\psi = \frac{P(x)}{r(x)^n}$ with $P(x)$ any (finite order) polynomial and 
$n \geq 1$ we obtain a finite norm element of $L^2$. If we take for 
instance $n=1$ then
\beq
\langle\psi,A\psi\rangle =-4\pi \int^{c}_{0}dxP(x)\left[\frac{d^2 P(x)}{dx^2}+
\mu_{0}P(x) \left(\alpha^2-\frac{2M}{r^3}-2\lambda\right)\right]
\feq  

thus, if $\alpha^2 > \frac{2M}{r_H^3}+2\lambda$, and we take $P$ to be any
polynomial on $x$ which is positive definite and has positive definite
second derivative in the interval $(0, c)$, then $ \langle\psi, A \psi\rangle \leq 0$
showing that the configuration is unstable.

\section{CONCLUSION}

We have carefully analyzed the reasons behind the possibility of scalar hair
in the asymptotically Anti de Sitter case comparing with the situation
in the asymptotically flat case. We have  discussed also the issue of
stability and found a very simple explanation which in fact points to the
instability within these theories and boundary conditions of the usual
Schwarzschild Anti de Sitter  solution. 
This work has dealt with the minimal coupled case, its   extension to 
the non-minimal coupled case is trivial  if we can  perform a conformal 
transformation  (i.e. if the required conformal factor  can be shown  to 
be nowhere vanishing) , in the nontrivial cases it is  hindered by the 
fact that in such case the control provided by the WEC over the signs of 
the various terms in eq. (\ref{teo2})  is lost.
We now  briefly note \cite{ADS/CFT} that according to the conjecture of ADS/CFT 
correspondence the Schwarzschild Anti de Sitter solution  of the theory in the bulk
should correspond to a thermal state of  the conformal theory in the  Anti
de Sitter ``boundary". But, as the black hole solution is unstable, so
should be the corresponding thermal state, and it seems very difficult to
envision what possibly could it be meant by a thermal state ( by
definition an  equilibrium state involving fluctuations) that  is unstable.
Needless is to say that such issues should be further investigated
and our point in mentioning them here is to note how the study of hairy
black holes can have implications in other, apparently disconnected
subjects.

\section{Acknowledgments}

D.S. wishes to thank A. Ashtekar,  A, Sen  and U. Nucamendi  for
helpful discussions. This work
was in part supported by DGAPA-UNAM Grant No. IN 112401 and by
CONACyT grant 32272-E. J.A.G. acknowledges support by CONACyT PhD. fellowship 149945.
D.S acknowledges partial support by the
Eberly Endowment and thanks  Penn State University for its hospitality.


\begin{thebibliography}{99}

\bibitem{EYM}P. Bizon, Phys. Rev Lett., {\bf 64} 2844 (1990);
M. S. Volkov and D. V. Gal'tsov, Sov. J. Nucl. Phys., {\bf 51} 1171 (1990);
H. P. Kunzle and A. K. M. Masood-ul-Alam, J. Math. Phys., {\bf 31}, 928
(1990).

\bibitem{ES}P. Bizon and T. Chamj, Phys. Lett. {\bf B 297}, 55,
(1992); M. Heusler, S. Droz, and N. Straumann, Phys. Lett. {\bf B268}, 371,
(1991);
{\bf B271}, 61, (1991); {\bf B258}, 21, (1992);

\bibitem{EYMH}B. R. Greene, S. D. Mathur, C. M. O'Neill,
 Phys. Rev. {\bf D 47}, 2242 (1993).

\bibitem{EYMd}
G. Lavrelashvili and D. Maison, Nucl. Phys. {\bf B 410}, 407 (1993).

\bibitem{Bek0} J.D. Bekenstein. {\it Ann. Phys. (NY)}, {\bf 82}: 535, (1974)

\bibitem{Heusler} M. Heusler, J. Math. Phys. 33, 3497, (1992)

\bibitem{EH}
D. Sudarsky,
 {\it Class. Quantum Grav.}, {\bf 12}, 579 (1995).

\bibitem{bekbh} J.D.Bekenstein Ann.Phys.(N.Y) 82, 535 (1974);
N. Bocharova, K. Bronikov and V. Melnikov, Vestn. Mosk. Univ.
 Fiz. Astron. 6, 706, (1970).

\bibitem{XZ} B. C. Xanthopoulos and T. Zannias J.Math.Phys.32,1875,(1991)

\bibitem{sudzann}
D. Sudarsky and T. Zannias. Phys. Rev. D {\bf 58}, 087502-1 (1998).

\bibitem{Zannias} T. Zannias J. Math. Phys. 36, 6970 (1995)

\bibitem{Saa}
A. Saa.
{\it J. Math. Phys.}, {\bf 37}, 2346,(1996).

\bibitem{Bekmayo}
A. E. Mayo and J. D. Bekenstein, Phys. Rev. D {\bf 54}, 5059 (1996).


\bibitem{Igor}
I. Pe\~na and D. Sudarsky. {\it Class. Quantum Grav}, {\bf14}, 3131 (1997)

\bibitem{Min} T. Torii, K. Maeda, and M.  Narita, Phys. Rev. {\bf D 64},
044007 (2001).

\bibitem{Nonmin} E. Wisttanley, gr-qc/ 0205092

\bibitem{us} D. N\'u\~nez, H. Quevedo, and D. Sudarsky,
 Phys. Rev. Lett. {\bf 76}, 571 (1996).

\bibitem{Sud-Wald} D. Sudarsky and R.M. Wald, Phys. Rev. D {\bf 46}, 1453
(1992).

\bibitem{unstableRN} K. Lee, V. P. Nair and E. Weinberg,
 Phys. Rev. Lett.
 {\bf 68}, 1100 (1992).

\bibitem{ADS/CFT}  A. Ashtekar and  A. Sen, private Communication

\bibitem{Mass} P. Chr\`usciel and W. Simon, gr-qc/0004032.

\bibitem{Inestability} R. M. Wald, J. Math. Phys. {\bf 33}, 1 (1992).

\end{thebibliography}
\end{document}